# Simulation-Based Cyber Data Collection Efficacy

David Thaw
School of Computing and Information / School of Law
University of Pittsburgh
Pittsburgh, USA
dbthaw@gmail.com

Bret Barkley
Department of Computer Science
University of Pittsburgh
Pittsburgh, USA
brb171@pitt.edu

Gerry Bella, Jr.
School of Computing and Information
University of Pittsburgh
Pittsburgh, USA
gdb17@pitt.edu

Carrie Gardner
Software Engineering Institute
Carnegie Mellon University
Pittsburgh, USA
cgardner@sei.cmu.edu

*Abstract*—Building upon previous research in honeynets and simulations, we present efforts from a two-and-a-half-year study using a representative simulation to collect cybersecurity data. Unlike traditional honeypots or honeynets, our experiment utilizes a full-scale operational network to model a small business environment. The simulation uses default security configurations to defend the network, testing the assumption that given standard security baseline, devices networked to the public Internet will necessarily be hacked. Given network activity appropriate for its context, results support the conclusion that no actors where able to break in, despite only default security settings.

## I. Introduction

It is a known problem that cybersecurity is lacking in comparative or experimental data, due to difficulties with data collection methodologies and concerns with sharing potentially sensitive data [1] [2]. This is particularly problematic for the state of cybersecurity, as decision makers set policies and procedures that are not always grounded in evidence from empirical research.

We contribute to the cybersecurity data collection effort by the development and deployment of representative simulations. These convincing simulations monitor and collect activity from mock users and potential intruders, generating high-fidelity data on how adversaries compromise systems and interact with them once inside.

The simulated network is tailored to reflect the perceived size and technical sophistication of the business or institution that the intruders believe they are interacting with. From the outside, it should appear as though a real organization is using the network to conduct its daily operations. This method is distinct from traditional honeypots or honeynets as discussed in [3]. Our method provides the model for studying our initial research hypothesis presented in this piece.

We theorized that deploying a representative simulation will not necessarily result in system compromise. This finding is in contrast to some conventional wisdom that by default all Internet-connected systems will be successfully hacked. We find no unexpected activity, malicious or otherwise anomalous, suggesting that no intruders were able to break in. Additionally, despite ongoing probing including known port scans and web crawlers, we did not find evidence of any further probing attempts such as targeted phishing emails to the simulated email addresses.

The rest of the paper is organized as follows. First, we review existing related research to honeypots, honeynets, and simulations and cyber data collection efforts. Second, we describe our research method and to frame and contextualize our experiment. Third, we present findings from the two-and-a-half-year study, yielding evidence which challenges the popular myth that "everything connected to the Internet necessarily will be hacked".

## II. Background

Traditional honeynets have a number of limitations in their ability to produce the types of empirical results discussed here. First, the honeypots they are comprised of generally employ limited (if any) camouflage and thus are easily identifiable as honeypots by even low- to moderately-sophisticated adversaries [4]. Second, deploying and reconfiguring honeynets for experimental purposes has been cost-prohibitive due to high costs of fixed-asset honeypots and a lack of utilization of virtualization technologies. Third, traditional honeypots -- even so-called "high interaction honeypots" -- generally have not employed active user behavior emulation methods.

These factors limit the ability of traditional honeynets to monitor adversaries' Tactics, Techniques and Procedures (TTPs) which depend on some type of user interaction to trigger an event [4]. However, advances in virtualization and cloud computing technologies have provided an alternative that allows for the development of comprehensive, flexible, and well-camouflaged simulations which can accurately emulate live production systems in an empirically-valid manner.

Table 1: Hallmark Honeypot, Honeynet, and Simulation Research

| Category | Features | Examples | References |
|---|---|---|---|
| Honeypot – Low Interaction | Basic listening services to log connections | HoneyC | [5], [6] |
| Honeypot – High Interaction | Services such as web servers to provide some sort of response | Argos | [7] [8] |
| Simulation | Full-scale environments with standard security configuration | Chameleon | [3] [9] |

Given the lack of robust methods to eliciting attackers to monitor their activity, we developed the Chameleon model to generate representative simulations to model, simulate, and collect dependable cybersecurity data [3]. We employ this model in this study, describing the simulation context in the next section. In this paper we present evidence of the feasibility of using Chameleon method, finding support that this method can attract activity, from which can monitor potential adversaries.

## III. METHODS AND MATERIALS

Building on our previous work, we instantiated a small financial services simulation [3]. This simulation sought to elicit attackers interested in financial data or targeting small business. We publicized the simulation with an alluring website to promote the company's services. Figure 1 below presents the symbolic Model-Object relationship for creating an instance of a model simulation geared towards a specific operational context.

**Figure 1: The Model-Object Relationship for our Simulations**

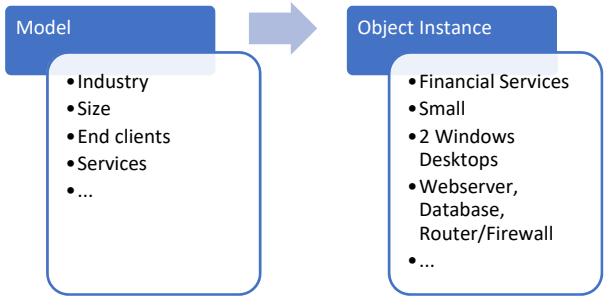

### A. Control Variables: Instance Features

We deployed the business with five systems: two windows desktops, a webserver, a customer database, and a router/firewall. In each system, we created artifacts appropriate to its operating context (for instance, realistically fictitious names, addresses, and social security numbers in the customer database). As an initial experiment, we intended the following to be evident: 1) the simulation should appear to be a small operation with minimal technical expertise implying minimal adherence to security best practices; 2) appliances and workstations configured using default settings from the manufacturer with little modification; 3) the database containing customer data should appear realistic; however, the corresponding social security numbers would not match an individual's birth time; 4) the simulation should consist of commonly used operating systems; 5) machines should be reasonably patched; and 6) should an attacker penetrate the firewall the machines contained within would appear to be workstations of the aforementioned small business [10].

The simulation used then-current versions of Windows, a commonly-used open-source software firewall installed on a commonly-used version of a UNIX-based operating system, and then-current versions of a common web server, all of which were maintained as virtualized appliances within our simulation environment. The network configuration included standard network address translation (NAT) with open outbound rules and no inbound permissions except port 80 (HTTP). Intra-simulation traffic was unrestricted. To maintain operational camouflage, we used a commercial Internet service provider, employing the level of service appropriate to the context of a business like that we simulated. This provided both general network camouflage and IP address camouflage.

### B. Experimental Variable: Security Configuration

The simulation was designed to minimize the number of attack vectors both to test hypothesis (2) and to create a "control" for future experiments. To accomplish this, the firewall configuration was restricted as described above, and no software or hardware with unpatched known vulnerabilities was deployed.

To further facilitate camouflage, the website was intentionally designed to appear outdated and was purposely designed to give attackers the impression that the business lacked technical sophistication.

Data points were collected using an array of separate collectors enabled on each machine or application. The collectors on the individual workstations would collect system logs generated by the operating system and forward them to a separate server, detached from the simulation. This data would be aggregated with logs from the web server and intrusion detection system (IDS) package from the firewall. These logs will correlate and validate abnormal activity detected within the simulation.

Prior to deploying our simulation on a live public network, we conducted a penetration test as a control to evaluate the any potential vulnerabilities and identify possible attack vectors we expect attackers to utilize. The results of that penetration test revealed no known vulnerabilities.

## IV. FINDINGS AND DISCUSSION

Our study had two experimental conditions: (1) determine if we could deploy camouflage adequate to attract representative potentially-malicious traffic; and (2) assuming (1), determine the extent to which a default baseline security configuration would be compromised over different time periods.

Regarding hypothesis (2), we specifically were interested in creating a baseline for evaluation of network penetration/attack surface risk. Many security professionals, scientists, and even the U.S. Department of Defense (DoD) have taken the position that any device connected to the public Internet essentially *will* be compromised, and the only question is *when*. While potentially true from a strict theoretical standpoint, such advice is less helpful to organizations which (unlike the DoD) lack the resources to maintain high-reliability operational environments. A more plausible hypothesis, we propose, firewall is that overinvestment in high-reliability perimeter defenses detracts from investment in other areas of security, creating more viable avenues of attack.

We evaluate this hypothesis using the simulation described above in Methods and Materials. We evaluated this configuration over approximately a two-year period examining both hypotheses (1) and (2). The dependent variables, presented in further detail below, are a series of known indicators-of-compromise (IoC) identified and measured on the emulated devices.

### A. Data

Our study ran from December 2015 - July 2018. During the course of the simulation, we did not inject any emulated user activity. We relied strictly upon capturing activity associated with external actors.

*B. Eliciting Activity*

Hypothesis (1) is important because, if our simulation fails to attract traffic representative of a what a contextually-similar organization would attract, our results will lack external and construct validity. Accordingly, we evaluate hypothesis (1) to determine if our simulation attracts such traffic. HTTP-based web traffic is our primary metric as this is the only open port and the primary method of "advertisement" of our simulation.

Our results, presented in the figures below, indicate that representative traffic was achieved for the period in question.

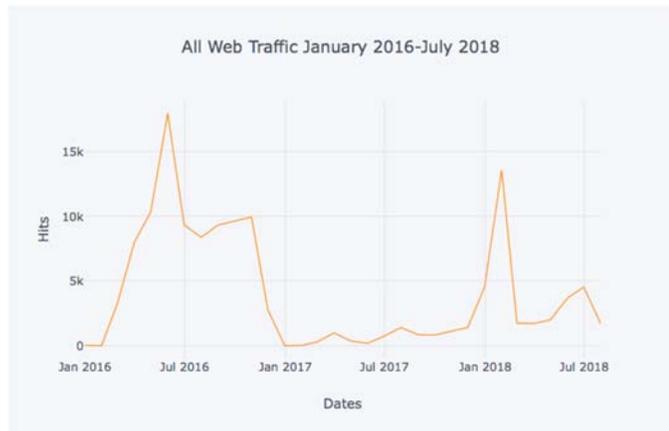

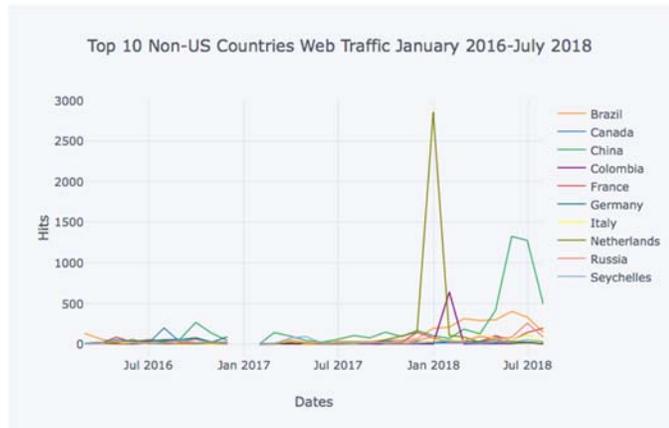

*C. (Attempted) Attacks?*

The simulation we designed accurately simulates the network of a small financial services firm. This is supported by our traffic data, which is appropriate for such an operation. Our data also shows evidence of potential adversaries gathering information about our network, through port scanning.

Despite many port scans and web crawls, our dataset finds no evidence of a compromise. We analyzed records from OS, application, and webserver logs and found zero indicators of compromise. This finding challenges the general attitude that simply being connected to the internet is a massive risk.

## V. LIMITATIONS & FUTURE WORK

Given the level of sophistication within the simulation and the lack of advertising outside of an active website, it is possible the simulation did not receive a significant level, or reasonably malicious variety, of traffic to gather enough data. While advertising methods are being evaluated in future simulations, we view this simulation as a network "idling" on the Internet. It is not making active requests for data, but it is available for requests and queries.

Additionally, we recognize that the simulation is not necessarily representative of all or even a portion of networks on the Internet. This project is a prototype experiment to test the feasibility of such an effort to realistically model and cole.

Ideally, the project's future simulations will focus on enhancing user simulation, and the programmatic development and deployment of custom simulations. We claim that the future of cybersecurity data collection will heavily rely on advanced simulation technologies. We aim for this project to serve as a prototype of such efforts.

ACKNOWLEDGMENT

We would also like to thank Chris Tomei, Chris Seifried, Jaclyn Ramsey, Ryan Craig, and Chris Lyons for their contributions to the project.